\documentclass[english]{article}
\usepackage[T1]{fontenc}
\usepackage[latin1]{inputenc}
\usepackage{amsmath}
\usepackage{graphicx}
\usepackage{setspace}
\usepackage{amssymb}

\makeatletter
\usepackage{babel}
\makeatother
\begin{document}
\begin{flushright}ITP-Budapest 614\end{flushright}

\begin{flushright}hep-ph/0410102\end{flushright}
\vspace{1.5cm}

\begin{center}{\LARGE Note on Unitarity Constraints in a Model for
a Singlet Scalar Dark Matter Candidate}\end{center}{\LARGE \par}
\vspace{1cm}

\begin{center}{\large G. Cynolter$^{*}$, E. Lendvai$^{*}$ and G.
Pócsik$^{\dagger}$ }\end{center}{\large \par}

\begin{onehalfspace}
\begin{center}\textit{$^{*}$Theoretical Physics Research Group of
Hungarian Academy of Sciences, Eötvös University, Budapest, Hungary}\\
\textit{$^{\dagger}$Institute for Theoretical Physics, Eötvös Lorand
University, Budapest, Hungary }\end{center}
\smallskip{}
\end{onehalfspace}

\begin{abstract}
We investigate perturbative unitarity constraints in a model \textit{}for
a singlet scalar dark matter candidate. \textit{}Considering elastic
two particle scattering processes of the Higgs particle and the dark
matter candidate, a real Klein-Gordon scalar field, perturbative unitarity
constrains the self-couplings of the scalar fields.
\end{abstract}
Despite the unique success of the Standard Model (SM) in the particle
physics accelerator experiments, there have been strong indications
of fundamental physics related to particle physics beyond the SM in
the past few years, such as neutrino oscillations and masses, the
baryon asymmetry of the universe cannot be explained by the SM, recent
fits to the cosmological parameters \cite{wmap} need dark energy.
About 23\% of the total energy density of the universe is made up
of some dark matter. Assuming that gravity does not change significantly
at distances larger than a few $kpc$ the dark matter must be non-baryonic
to maintain the success of big bang nucleosynthesis.

These problems were addressed among others in \cite{murayama} where
a possible minimal extension of the SM was proposed. They added the
minimal number (6) of new degrees of freedom purely to answer the
empirical challenges. The non-baryonic dark matter candidate is assumed
to be a $Z_{2}$ symmetric gauge singlet scalar field, $S$. It can
account for the observed dark matter abundance and is consistent with
the limit from CDMS-II experiment \cite{cdms}. \cite{murayama} considered
a few consequences of the model such as triviality and stability of
the Higgs potential, Higgs decays into new particles. 

In this note we apply perturbative unitarity \cite{lee} to a singlet
dark matter field $S$ coupled to the Higgs field $H$ and itself
in the SM completed by terms describing $S,H$ interactions. The constraints
are valid also for \cite{murayama} because the inflaton field of
\cite{murayama} is too heavy to participate in the scattering. We
get various upper bounds for the relevant scalar couplings.

Start with the minimal renormalizable extension of the Standard Model
providing a scalar non-baryonic dark matter candidate $S$. $S$ is
required to be odd under a $Z_{2}$ symmetry in order to be stable.
The odd singlet scalar can have renormalizable interaction only with
the standard Higgs and not with the ordinary fermions and gauge bosons.
This model was proposed earlier in the literature by Silveira and
Zee \cite{zee}, a complex scalar field case was studied in \cite{mcdonald}.
The parameters of the model were first constrained by Burgess \textit{et
al.} \cite{pospelov}, later in \cite{murayama}.

The Lagrangian of the scalar sector is \begin{equation}
L_{SH}=\left|D_{\mu}H\right|^{2}-\frac{\lambda}{2}\left|H^{\dagger}H-\frac{v^{2}}{2}\right|+\frac{1}{2}\partial_{\mu}S\partial^{\mu}S-\frac{1}{2}m_{0}^{2}S^{2}-\frac{k}{2}H^{\dagger}H\, S^{2}-\frac{{\lambda_{S}}}{4!}S^{4}.\label{eq:L}\end{equation}
Beside the usual Higgs potential parameters $\lambda,$$v=254$ GeV,
there are three new parameters $m_{0},k,\lambda_{S}$ determining
the properties of $S$. The potential of the H-S sector is bounded
from below if $\lambda,\lambda_{S}>0$ and $k>0$ or\begin{equation}
3k^{2}<\lambda_{S}\lambda\qquad\textrm{for }k<0.\label{eq:2}\end{equation}
 The Higgs field gets a vacuum expectation value $v$ while $\left\langle S\right\rangle =0$
in order to respect the $Z_{2}$ symmetry. The Higgs mechanism generates
a mass of $m_{H}^{2}=\lambda v^{2}$ for the Higgs and also contributes
to the mass of the $S$ particle\begin{equation}
m_{S}^{2}=m_{0}^{2}+\frac{1}{2}kv^{2}.\label{eq:3}\end{equation}
$m_{s}^{2}>0$ is required for $\left\langle H\right\rangle =(0,v/\sqrt{2})$
and $\left\langle S\right\rangle =0$ be a local minimum. This is
also a global minimum as long as $m_{0}^{2}>-\frac{1}{2}v^{2}\sqrt{\frac{1}{3}\lambda\lambda_{S}}$
\cite{pospelov}. After electroweak symmetry breaking in the unitarity
gauge the potential of the scalar sector becomes\begin{equation}
V_{SH}=\frac{\lambda}{8}H^{4}+\frac{\lambda}{2}H^{3}v+\frac{\lambda}{2}H^{2}v^{2}+\frac{1}{2}\left(m_{0}^{2}+\frac{1}{2}kv^{2}\right)S^{2}+\frac{k}{2}vHS^{2}+\frac{k}{4}H^{2}S^{2}+\frac{\lambda_{S}}{4!}S^{4}.\label{eq:VhS}\end{equation}

Next we apply tree-level perturbative unitarity \cite{lee} to scalar
elastic scattering processes in the model (\ref{eq:VhS}). The zeroth
partial wave amplitude, \begin{equation}
a_{0}=\frac{1}{32\pi}\sqrt{\frac{4p_{f}^{CM}p_{i}^{CM}}{s}}\int_{-1}^{+1}T_{2\rightarrow2}d\cos\!\theta\label{eq:partw}\end{equation}
 must satisfy $\left|Rea_{0}\right|\leq\frac{1}{2}$. $s$ is the
centre of mass (CM) energy and $p_{i,f}^{CM}$are the inital and final
momenta in CM system.

There are three possible two particle states $HH,HS,SS$ and four
scattering processes. Inclusion of the gauge bosons does not significantly
alter our consideration since they do not interact with the new singlet
scalar $S$ at tree level.

1.) $HH\rightarrow HH$. The tree-graphs contributing to this process
are drawn in Fig. 1. 

\begin{figure}[h]
\includegraphics[%
  scale=0.8]{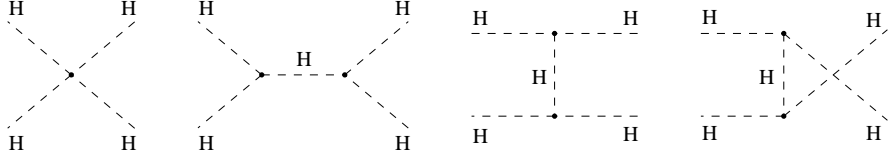}

\caption{Tree level Feynman diagrams for HH$\rightarrow$HH in the SM}
\end{figure}
 We get\begin{equation}
T_{HH\rightarrow HH}=3\frac{m_{H}^{2}}{v^{2}}\left(1+3m_{H}^{2}\left(\frac{1}{s-m_{H}^{2}}+\frac{1}{t-m_{H}^{2}}+\frac{1}{u-m_{H}^{2}}\right)\right).\label{eq:thh}\end{equation}

2.) $SS\rightarrow SS$. The contact graph and the H-exchange graphs
can be seen in Fig. 2.

\begin{figure}[h]
\includegraphics[%
  scale=0.8]{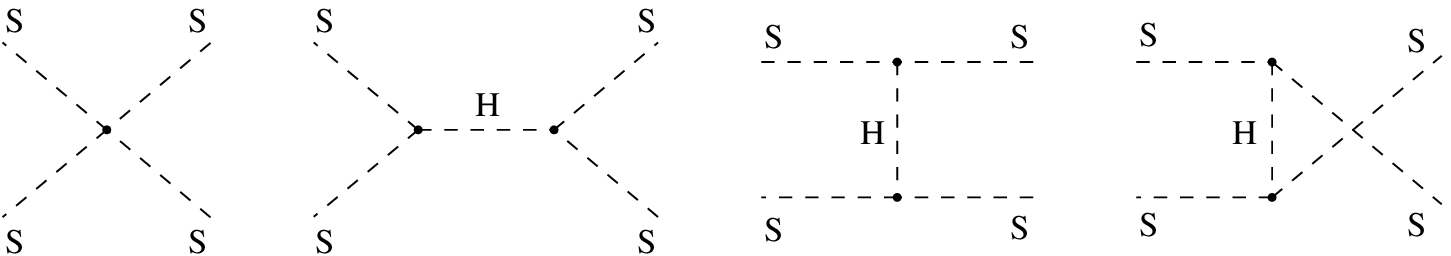}

\caption{Tree level Feynman diagrams for SS$\rightarrow$SS }
\end{figure}

\begin{equation}
T_{SS\rightarrow SS}=\lambda_{S}+k\left(\frac{kv^{2}}{s-m_{H}^{2}}+\frac{kv^{2}}{t-m_{H}^{2}}+\frac{kv^{2}}{u-m_{H}^{2}}\right).\label{eq:tss}\end{equation}

3.) $SS\rightarrow HH$. 

\begin{figure}[h]
\includegraphics[%
  scale=0.8]{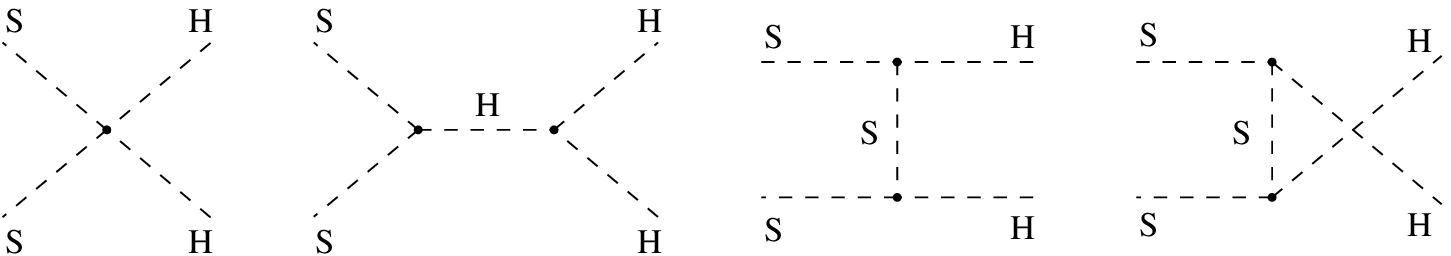}

\caption{Tree level Feynman diagrams for SS$\rightarrow$HH}
\end{figure}
 The T matrix (Fig. 3) is 

\begin{equation}
T_{SS\rightarrow HH}=k\left(1+3m_{H}^{2}\frac{1}{s-m_{H}^{2}}+kv^{2}\left(\frac{1}{t-m_{S}^{2}}+\frac{1}{u-m_{S}^{2}}\right)\right).\label{eq:tsshh}\end{equation}

4.) $HS\rightarrow HS$. The T matrix (Fig. 4.) is 

\begin{equation}
T_{HS\rightarrow HS}=k\left(1+v^{2}\left(\frac{k}{s-m_{S}^{2}}+\frac{3\lambda}{t-m_{H}^{2}}+\frac{k}{u-m_{S}^{2}}\right)\right).\label{eq:tsh}\end{equation}

\begin{figure}[h]
\includegraphics[%
  scale=0.8]{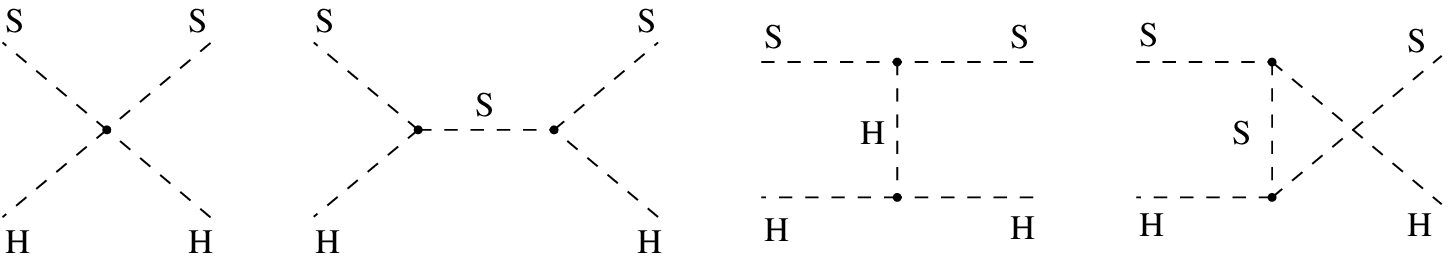}

\caption{Tree level Feynman diagrams for SH$\rightarrow$SH}
\end{figure}

From (\ref{eq:partw}) we have the following partial wave projection
of the coupled system in the J=0 channel:\begin{eqnarray}
a_{0}(HH\rightarrow HH) & = & \frac{3m_{H}^{2}}{16\pi v^{2}}\sqrt{1-\frac{4m_{H}^{2}}{s}}\left(1+\frac{3m_{H}^{2}}{s-m_{H}^{2}}-\frac{6m_{H}^{2}}{s-4m_{H}^{2}}\ln\left(\frac{s}{m_{H}^{2}}-3\right)\right),\nonumber \\
a_{0}(\: SS\,\rightarrow\, SS\,) & = & \frac{1}{16\pi}\sqrt{1\!-\!\frac{4m_{S}^{2}}{s}}\left(\lambda_{S}+\!\frac{k^{2}v^{2}}{s-m_{H}^{2}}\!-\frac{2k^{2}v^{2}}{s-4m_{S}^{2}}\ln\left(\frac{s-4m_{S}^{2}}{m_{H}^{2}}\!+1\!\right)\!\right),\nonumber \\
a_{0}(SS\rightarrow HH) & = & \frac{k}{16\pi}\kappa_{SS\rightarrow HH}\left(1+3m_{H}^{2}\left(\frac{1}{s-m_{H}^{2}}\right)-\frac{2kv^{2}}{\sqrt{s-4m_{S}^{2}}\sqrt{s-4m_{H}^{2}}}\right.\nonumber \\
 &  & \left.\ln\left(1+\frac{2\sqrt{s-4m_{S}^{2}}\sqrt{s-4m_{H}^{2}}}{s-\sqrt{s-4m_{S}^{2}}\sqrt{s-4m_{H}^{2}}-2m_{h}^{2}}\right)\right),\label{eq:a0}\\
a_{0}(SH\rightarrow SH) & = & \frac{k}{16\pi}\kappa_{SH\rightarrow SH}\left(1+\frac{kv^{2}}{s-m_{H}^{2}}-\frac{3s}{A}m_{H}^{2}\ln\frac{s-3m_{H}^{2}}{s\left(2m_{S}^{2}-m_{H}^{2}\right)-\left(m_{H}^{2}-m_{S}^{2}\right)^{2}}-\right.\nonumber \\
 &  & \left.-\frac{s}{A}kv^{2}\ln\frac{\left(s-3m_{S}^{2}\right)s+\left(m_{H}^{2}-m_{S}^{2}\right)^{2}}{s\left(2m_{H}^{2}-m_{S}^{2}\right)}\right),\nonumber \end{eqnarray}
where $A=s^{2}-2s\left(m_{H}^{2}+m_{S}^{2}\right)+\left(m_{H}^{2}-m_{s}^{2}\right)^{2}$,
$\kappa_{SS->HH}=\left(1-\frac{4m_{S}^{2}}{s}\right)^{1/4}\left(1-\frac{4m_{H}^{2}}{s}\right)^{1/4}$
and $\kappa_{SH->SH}=\sqrt[]{\left(1-\frac{(m_{H}+m_{S})^{2}}{s}\right)\left(1-\frac{(m_{H}-m_{S})^{2}}{s}\right)}$
are kinematical factors.

Perturbative unitarity will be imposed in the high energy limit $s\gg m_{H}^{2},m_{S}^{2}.$
The coupled amplitudes are \[
\left(\begin{array}{ccc}
a_{0}^{HH\rightarrow HH} & a_{0}^{HH\rightarrow SS} & a_{0}^{HH\rightarrow SH}\\
a_{0}^{SS\rightarrow HH} & a_{0}^{SS\rightarrow SS} & a_{0}^{SS\rightarrow SH}\\
a_{0}^{SH\rightarrow HH} & a_{0}^{SH\rightarrow SS} & a_{0}^{SH\rightarrow SH}\end{array}\right)\begin{array}{c}
\\\longrightarrow\\
{\scriptstyle s\gg m_{H}^{2},m_{S}^{2}}\end{array}\frac{1}{16\pi}\left(\begin{array}{ccc}
3\lambda & k & 0\\
k & \lambda_{S} & 0\\
0 & 0 & k\end{array}\right).\]

Requiring $\left|Rea_{0}\right|\leq\frac{1}{2}$ for each individual
process above we obtain \begin{eqnarray}
HH\rightarrow HH &  & m_{H}\leq\sqrt{\frac{8\pi}{3}}v.\label{eq:Hunit1}\\
HS\rightarrow HS\textrm{ and }HH\rightarrow SS &  & |k|\leq8\pi.\label{eq:k8pi}\\
SS\rightarrow SS &  & \lambda_{S}\leq8\pi.\label{eq:lambda8pi}\end{eqnarray}

While (\ref{eq:Hunit1}) is the well known bound in \cite{lee}, (\ref{eq:k8pi})
shows that the maximum contribution of the Higgs mechanism to $m_{S}$
is 900 GeV.

(\ref{eq:k8pi}) can be improved for $k<0$ by using (\ref{eq:Hunit1},\ref{eq:lambda8pi})
in the positivity relation \begin{equation}
-k\leq\frac{8\pi}{3}\sim4.2\;\textrm{and }\frac{1}{2}kv^{2}\gtrsim-(520\textrm{ GeV})^{2}.\label{eq:kcon}\end{equation}
For $m_{H}=150$GeV, $\lambda=0.38$ this goes into $\frac{1}{2}kv^{2}\gtrsim-(230\textrm{ GeV})^{2}$.

These bounds can be refined considering the partial wave unitarity
for the three coupled channel system HH, SS, SH and constraining the
eigenchannel with the highest eigenvalue. Actually SH decouples from
HH and SS and the remaining two eigenvalues are $\mu_{1,2}=\frac{3\lambda+\lambda_{S}}{2}\pm\frac{1}{2}\sqrt{\left(3\lambda+\lambda_{S}\right)^{2}+4k^{2}},$
providing the constraint \begin{equation}
\lambda_{S}+\frac{k^{2}}{8\pi}\leq8\pi-3\lambda.\label{eq:unitmix}\end{equation}
The constraint (\ref{eq:unitmix}) contains all the previous bounds
(\ref{eq:Hunit1}, \ref{eq:k8pi}, \ref{eq:lambda8pi}).

After the new measurement of the top mass \cite{d0} the upper bound
of the Higgs mass from radiative corrections is 251 GeV at 95 \% C.L.
and the direct lower bound is 114.5 GeV from LEP2 implying the range
$0.21-0.97$ for $\lambda$. We see, however, that for $\lambda=0.21-0.97$
($m_{H}=114.5$GeV-251 GeV) the right hand side of (\ref{eq:unitmix})
changes very small, 24.5-22.1, and $\lambda_{S}+\frac{k^{2}}{8\pi}\lesssim8\pi.$
Only a heavy Higgs would provide a stronger upper bound.

In conclusion we have considered a simple non-baryonic dark matter
candidate model added to the Standard Model. We have calculated the
J=0 partial wave amplitudes for the two particle elastic scattering
processes and found that the scalar couplings of the model are restricted.

In general our results did not restrict the $S$ mass, however, assuming
$m_{S}$ comes from the Higgs mechanism we get $m_{S}<900$GeV. The
model can account for the dark matter in the universe and is consistent
with the limits from CDMS-II. Perturbative unitarity constraints allow
also higher $\lambda_{S},k$ than those obtained from stability and
triviality described in \cite{murayama}.


\newpage

\begin{thebibliography}{1}
\bibitem{wmap}D.N. Spergel \textsl{et al.} {[}WMAP Collaboration{]}, Astrophys.
J. Suppl. \textbf{148} (2003) 175.
\bibitem{murayama}H.Davoudiasl, R.Kitano, T.Li and H.Murayama, arXiv:hep-ph/0405097.
\bibitem{cdms}D.S.Akerib \textit{et al.} {[}CDMS Collaboration{]}, arXiv:astro-ph/0405033
\bibitem{lee}B.W. Lee, C. Quigg and H.B.Thacker, Phys. Rev. D \textbf{16} (1977)
1519.
\bibitem{zee}V.Silveira and A.Zee, Phys. Lett. B \textbf{161} (1985) 136.
\bibitem{mcdonald}J.McDonald, Phys. Rev. D \textbf{50} (1994) 3637.
\bibitem{pospelov}C.P.Burgess, M.Pospelov and T.ter Veldhuis, Nucl. Phys. B\textbf{619}
(2001) 709 {[}arXiv:hep-ph/0011335{]}.
\bibitem{d0}DØ Collaboration, Nature \textbf{429}, 638 (2004)\end{thebibliography}
\end{document}